# MOFSimplify: Machine Learning Models with Extracted Stability Data of Three Thousand Metal–Organic Frameworks

Aditya Nandy[1,2,#], Gianmarco Terrones[1,#], Naveen Arunachalam[1], Chenru Duan[1,2], David W. Kastner[1,3], and Heather J. Kulik[1,*]

[1]*Department of Chemical Engineering, Massachusetts Institute of Technology, Cambridge, MA 02139*

[2]*Department of Chemistry, Massachusetts Institute of Technology, Cambridge, MA 02139*

[3]*Department of Biological Engineering, Massachusetts Institute of Technology, Cambridge, MA 02139*

[#]These authors contributed equally.

ABSTRACT: We report a workflow and the output of a natural language processing (NLP)-based procedure to mine the extant metal–organic framework (MOF) literature describing structurally characterized MOFs and their solvent removal and thermal stabilities. We obtain over 2,000 solvent removal stability measures from text mining and 3,000 thermal decomposition temperatures from thermogravimetric analysis data. We assess the validity of our NLP methods and the accuracy of our extracted data by comparing to a hand-labeled subset. Machine learning (ML, i.e. artificial neural network) models trained on this data using graph- and pore-geometry-based representations enable prediction of stability on new MOFs with quantified uncertainty. Our web interface, MOFSimplify, provides users access to our curated data and enables them to harness that data for predictions on new MOFs. MOFSimplify also encourages community feedback on existing data and on ML model predictions for community-based active learning for improved MOF stability models.

## 1. Background and Summary

Metal–organic frameworks (MOFs) have reticular chemistry and well-defined, isolated metal sites[1,2] that make them promising for applications in gas adsorption[3,4], sensing[5,6], separations[7,8], and catalysis[9-14]. The modular nature of MOFs enables the design of hypothetical materials libraries amenable to virtual high throughput screening (VHTS) by combining distinct organic linkers, inorganic building blocks, and topologies to make a MOF.[2,7,15-19] The number of experimentally realizable MOFs with varying metals, linkers, and pore size has grown rapidly[20], despite challenges in synthesis[21,22] and post-synthetic modification[23]. After synthesis, MOFs must also undergo activation (i.e., solvent removal from pores) to enable their practical use. Despite advances in experimental methods[24,25] for MOF activation, many MOFs are unstable upon activation[22,26,27] and thus unusable.[28] For practical use as catalysts or functional materials, MOFs must also sustain their porosity and structural integrity at elevated temperatures.[29-34]

VHTS efforts for screening hypothetical MOFs typically rely heavily on expert intuition for identifying candidate materials that are then synthesized.[2,7,35] Although heuristics such as pore size[36] or hard-soft acid base theory[37] for predicting metal–linker bond strength are frequently invoked to predict MOF stability, numerous exceptions exist, limiting the broad applicability of heuristics for stability prediction.[36,38-41] Rules for thermal stability derived from subsets of MOFs do not extrapolate well to new MOFs outside of those subsets.[42] Molecular mechanics models that are useful for VHTS with MOFs also cannot predict activation stability.[19,43-45]

Limitations in using computation[19,43,44] or heuristics[36,37] to predict stability motivates data driven machine learning (ML) models trained on large experimental data sets. Gaining experimental solvent removal and thermal stability data in sufficient quantities to train ML models, however, remains a formidable challenge. Although a few studies have gathered experimental data from a single source[42,46] to reveal stability trends, the unified efforts of thousands of researchers over multiple decades represents an untapped source of knowledge[47] for the factors that govern MOF stability. Natural language processing[48] (NLP) is a promising approach to leverage this data from the literature. Many studies have combined NLP of the extant literature with ML to identify synthesis conditions for inorganic materials.[49-51] NLP has been used to quantify the role of organic structure directing agents in governing zeolite topology.[52] However, a lack of systematic naming[53,54] in MOF chemistry (e.g., HKUST-1 and Cu-BTC are the same MOF) has limited the use of NLP-based named entity recognition for the design of new MOFs. While NLP has worked well for identifying MOF properties such as surface area through their unique units,[55,56] human interpretation of the structure name is required to relate extracted properties back to the original structure.[55,56] Due to challenges in mapping MOF names to structures[53,54], coupled with the lack of unique units or measurements for stability assessments, no efforts have collated data on MOF stability.

We recently leveraged[57] the extant literature to identify how MOF linker and inorganic secondary building unit (SBU) composition as well as MOF connectivity govern MOF stability. We utilized NLP to curate stability-related experimental properties for structurally characterized MOFs. From the curated data, we trained artificial neural network (ANN) models that achieve high prediction accuracies for solvent removal stability (accuracy: 0.76, area under the receiver operating curve: 0.79) and thermal stability (mean absolute error: 47°C). These models use revised autocorrelations[18,58] (RACs) as fingerprints for each MOF that are derived from the MOF's clean (e.g., without solvent or disorder) crystallographic information file (CIF). Models trained on this data revealed the importance of both linker and metal features, demonstrating why solely metal-based or linker-based heuristics fail to predict MOF stability.

Here, we tabulate data on solvent removal stability for 2,179 structures and thermal stability for 3,132 structures of MOFs reported in the experimental literature. Our data set is the first source to map MOF experimental stabilities to well-defined experimental structures. We also provide representative linkers and SBUs from each structure, and the fingerprints we used to construct our data-driven models. We demonstrate how users can utilize our data sets, make predictions on new materials, or improve the quality of labels for our experimental stability data set. Our dataset and methodology will enable the curation of more reports of MOF stability, paving the way for the design of stable MOFs.

## 2. Methods

*Data Mining*

The starting point for data curation was the all solvent removed portion of the 2019 Computation-ready, experimental (CoRE) MOF database v1.1.2, which contains 10,143 non-disordered structures.[59] Of this set, the 9,597 MOFs that were compatible with the generation of graph-based revised autocorrelation[18,58] (RAC) and geometric[60,61] features were retained for further filtering steps (see Data Records). A subset of 9,202 MOFs were sanitized[59] structures from the Cambridge Structural Database[62,63] (CSD) that could be associated with a unique CSD refcode. We used these refcodes to obtain the digital object identifier (DOI) of the manuscript associated with each structure in the CSD[62] v5.41, released in November 2019. In total, 8,809 refcodes had associated DOI entries, which corresponded to 5,152 unique manuscripts (Figure 1).

We used the ArticleDownloader[64] package to automatically obtain manuscripts from the Royal Society of Chemistry (RSC), Wiley-VCH, the American Association for the Advancement of Science (AAAS), Springer, and Nature. Articles from the American Chemical Society (ACS) were obtained via a direct download agreement between ACS and the Massachusetts Institute of Technology. Through this procedure, 3,809 manuscripts were downloaded, which corresponded to 7,004 structures, in HTML or XML format for subsequent text extraction and parsing (Figure 1). For the 1,343 manuscripts associated with 1,805 structures that could not be automatically downloaded in HTML or XML format, roughly two-thirds were only available as PDFs (938 manuscripts, 1,307 structures) and one-third could not be automatically downloaded (405 manuscripts, 498 structures).

Next, we used text parsing on our corpus to determine labels (i.e. unstable or stable) for the solvent removal stability of MOFs and to identify manuscripts that contain thermogravimetric analysis (TGA) data (Table 1). We tokenized all manuscripts into sentences using the ChemDataExtractor[65] package. Because information about MOF solvent removal and thermal stability does not appear in a specific experimental methods section (e.g., as is the case for synthesis), text search of the entire manuscript is necessary. We avoided false positives (i.e., from introductory text) by excluding sections labeled as introductions, and only analyzed the last 60% of the manuscript text for letters or communications that lacked section headers (Figure 1).

Complex sentence structure limited the utility of sentiment-based models (e.g., VADER[66] sentiment) in identifying stability. We employed syntactic dependency parsing to extract labels for MOF solvent removal stability. First, we pattern matched (i.e., used regular expressions) for keywords pertaining to common MOF solvents, MOF structural integrity, and the process of MOF activation, identifying a set of sentences relating to activation stability. We used additional regular expressions to eliminate sentences relating to air or water stability, or activation processes that are not MOF activation (e.g., catalytic C–H activation). Next, we performed dependency parsing using the Stanza[67] NLP toolkit. Through dependency parsing, we analyzed pairwise mappings of words and disambiguated negations that are challenging to distinguish with regular expressions (e.g., "no crystallinity" vs. "no loss of crystallinity") for the manuscripts containing relevant sentences (2,649 out of 3,809). We then assigned each sentence a label of unstable (0) or stable (1). Because most manuscripts report on more than one MOF, we only assigned labels to manuscripts where all sentences had the same label (1,209 out of 2,649 manuscripts, Figure 1). We then assigned all MOFs from a given labeled manuscript the text-mined manuscript label. Finally, we eliminated 111 MOFs that had identical connectivity (e.g., same RACs), but conflicting text-mined labels from different manuscripts. In total, we identified 2,179 labels for solvent removal stability corresponding to structures from the CoRE MOF 2019 dataset (Figure 1).

For thermal stability analysis, we performed regular expression searches to identify a subset of 2,366 manuscripts (out of the 3,809 downloadable manuscripts) that could be expected to contain a TGA trace (Table 1). Of this set, 1,886 contain one or more TGA traces corresponding to CoRE MOF 2019 structures. The remainder either lack a TGA trace or only contain TGA traces for structures not deposited in the CoRE MOF 2019 database. Because TGA decomposition temperatures ($T_d$) are reported in a number of ways that could refer to the onset temperature or temperature of complete collapse, we extracted all critical TGA trace temperatures following a consistent protocol using WebPlotDigitizer[68]. We obtained two lines representing the TGA data before and after decomposition from four points on the TGA trace and calculated the intersection point of the two lines to obtain $T_d$ (Figure 1). Overall, we identified 3,132 thermal decomposition temperatures that corresponded to featurized CoRE MOF 2019 structures.

*Building Blocks and Descriptors*

First, we obtained the primitive unit cell for each MOF in the CoRE MOF 2019 database using pymatgen.[69] We divided each MOF into its constituent inorganic secondary building units (SBUs) and organic linkers during the generation of RACs[58,70,71] using molSimplify[18] (Table 2). To identify unique SBUs and linkers in each MOF, we computed the atom-weighted molecular graph determinants[72] and obtained the relevant subgraphs in the MOF components with unique determinants (see Data Records). In addition, we computed geometric properties (e.g. maximum included sphere) with Zeo++ using a nitrogen probe molecule (Table 3).[60,61] All artificial neural network (ANN) models use RACs and geometric features as inputs to make predictions and were trained using keras[73] with a tensorflow[74] backend (see Data Records).

## 3. Data Records

We provide two JSON files, one for MOFs with solvent removal stability labels (solvent_removal_stability.json) and the other for MOFs with thermal stability labels (thermal_stability.json). The solvent removal stability JSON file contains 2,179 entries, and the thermal stability JSON file contains 3,132 entries.

Within the JSON files, each MOF structure is tabulated as a separate entry. In the solvent removal stability JSON file each entry contains the refcode of a MOF (i.e., as in the CoRE MOF 2019 database[59] and the CSD[62,63]), the DOI of the associated manuscript, sentences identified during regular expression matching and their corresponding locations in the manuscript, and the data partition for ANN usage[57] (e.g., train, validation, or test). Additionally, in each entry we report RAC[18,58] and geometric features[61]; ANN prediction probabilities, which are float values between 0 and 1, with values < 0.5 (≥ 0.5) corresponding to instability (stability) upon solvent removal, respectively; and ANN latent space entropy[75] measurements (which have a maximum value of 0.693 for binary classification) from training data. We also provide blocks for each unique inorganic SBU and organic linker in TRIPOS mol2 format, which can be automatically loaded into molSimplify[76] for structure manipulation. We determine whether or not each linker or SBU is unique by computing atom-weighted molecular graph determinants[72], and we keep only one representative example of each linker and SBU with a unique molecular graph determinant.

In addition to the entry information provided in the solvent removal stability JSON file, the thermal stability JSON file contains the four extracted points from the TGA trace of each MOF with a thermal stability label. We provide ANN predictions ($T_d$*) in units of degrees Celsius, and we provide latent space distance (i.e., both scaled and unscaled) measurements that can be used for uncertainty quantification[77] in regression models. The scaled latent space distances have a maximum distance of 1 with respect to the training data, in accordance with prior work[57].

As an alternative to the JSON files, we provide XLSX files for the solvent removal stability and thermal stability data sets. These XLSX files contain 2,179 and 3,132 entries respectively, and they contain the same information as the JSON files. We also provide TRIPOS mol2 files for the representative extracted inorganic SBUs and organic linkers separately.

We provide the refcodes, DOIs, and extracted sentences as an XLSX file for the structures for which we could identify keywords but could not assign a unique label. For solvent removal stability, multiple sentences may have different labels, preventing the assignment of an unambiguous final label (e.g., both positive and negative stability identified or challenging disambiguation of MOF structures). For thermal stability, TGA may be mentioned within the manuscript, but a TGA trace corresponding to the MOF in the CoRE MOF 2019 database may not be identifiable (e.g., when there are multiple structures corresponding to a manuscript).

Lastly, we provide our two ANN models (solvent removal stability classification and thermal stability regression) from prior work[57] as .h5 files that can be used with our open-source Python scripts, found on our GitHub repository (see Code Availability). We provide all JSON files, Excel sheets, SBU and linker structures, and models at our Zenodo repository 10.5281/zenodo.5508357.

On the MOFSimplify website (see Usage Notes), the user can download information on latent space nearest neighbor (LSNN) MOFs to a MOF input by the user. These LSNN MOFs are drawn from model training data, and the user can download information on them in the form of TXT, CIF, and CSV files. The TXT files each describe one MOF and include the experimentally observed stability for the MOF, the associated DOI, and its latent space distance to the MOF input by the user. In addition, LSNN CoRE MOF 2019 structures can be downloaded as CIF files. For thermal stability LSNN MOFs, the user can download their simplified TGA data as CSV files.

## 4. Technical Validation

We obtained a random sample of 100 MOFs from our solvent removal stability dataset to assess the quality of our NLP-assigned solvent removal stability labels in comparison to manual interpretation by a scientist. Over this set, only two cases of MOFs that are incorrectly labeled as unstable upon solvent removal but are stable upon solvent removal. The majority (i.e., 78 MOFs) are correctly labeled, 47 of which are stable and 31 unstable upon solvent removal (Figure 2). For the remaining 20 MOFs, the extracted sentences do not make a definitive statement about solvent removal stability, with 9 cases labeled as stable and 11 unstable (Figure 2). Analyzing the cases where the NLP workflow definitively assigns stability but the sentences are more ambiguous, these cases either mention another aspect of stability (e.g. stable coordination environment) while mentioning solvents, or mention that solvent removal stability was evaluated without stating the outcome.

To extract $T_d$ for thermal stability labels, we used NLP only to identify the presence of the TGA trace, which we then systematically digitized. Because thermal stability is not reported consistently across manuscripts (e.g., some manuscripts report decomposition onset temperatures, and others decomposition completion temperatures), we extracted $T_d$ from TGA traces consistently, using the start and the end of the decomposition step (see Methods). This process makes the thermal stability quantitative data less sensitive than solvent removal stability to either the NLP protocol or the method of reporting by the researcher.

As an example of the benefit of systematic extraction of $T_d$, we select a representative manuscript[78] (DOI: 10.1002/slct.201600844) that contains ten MOFs. Only six of these MOFs (SANGEW, SANGUM, SANHIB, SANHOH, SANHUN, and SANJAV) are present in the CoRE MOF 2019 dataset, whereas the remaining four MOFs (SANGIA, SANGOG, SANHAT, and SANHEX) are not. As a result, the latter four MOFs are excluded from our dataset. The manuscript reports all ten MOFs "remain thermally stable until 553K" and states that the first step of the TGA trace corresponds to the loss of a guest molecule, while the second step corresponds to decomposition. Our procedure uses the unit cell parameters provided in the supporting information to identify the CSD refcodes corresponding to the MOF labels in the manuscript (SANGEW: MOF1, SANGUM: MOF4, SANHIB: MOF7, SANHOH: MOF8, SANHUN: MOF9, SANJAV: MOF10) and then uses these name–structure mappings to associate a TGA trace with each MOF. The digitization procedure uses the beginning and end of the second step of each TGA trace to quantify decomposition temperatures. For SANHOH and SANHUN, manual inspection of the TGA trace reveals that decomposition starts near 300°C and ends near 400°C. In contrast, for SANGEW, SANGUM, SANHIB, and SANJAV, decomposition also starts near 300°C but does not conclude until 600°C (Figure 2). Although these MOFs begin decomposing at similar temperatures, all are below the value reported in the text by the authors, and some MOFs decompose more slowly than others. This case study demonstrates the differences in how TGA trace results are reported and quantified, motivating a systematic analysis and labeling. From the systematically labeled data, our final distribution of extracted $T_d$ values over the thermal stability dataset is a normal distribution centered around 358°C with a 87°C standard deviation (Figure 2).

As an additional blinded test, we hand-labeled the solvent removal stability and thermal stability of 40 MOFs from manuscripts that could not be automatically downloaded from the publisher (i.e., from Elsevier in this case). These data points are not present within the entire (i.e., train or test) solvent removal stability or thermal stability data sets. From these 40 MOFs, 20 were assigned stable and 20 unstable with respect to solvent removal by our solvent removal stability ANN. Over this hand-labeled set, we find that the ANN correctly predicted the stability of the majority (i.e., 31 out of 40) of this set of MOFs. For the remaining 9 MOFs, 7 stable MOFs were predicted to be unstable, while 2 unstable MOFs were predicted to be stable. This 78% accuracy is comparable to the ANN test set performance. We find that the mean absolute error (MAE) of the $T_d$ predictions generated by the thermal stability ANN on the hand-labeled MOFs is 55°C, which is comparable to the test set performance (MAE: 47°C) of the thermal stability ANN (Figure 2). The comparable performances on unseen data demonstrate that we can use our models to screen unseen MOFs to quantitatively predict their activation and thermal stabilities.

## 5. Usage Notes

We introduce the MOFSimplify website mofsimplify.mit.edu, a tool for analyzing and making property predictions on MOFs (Figure 3). To use MOFSimplify, the user selects a MOF for analysis in CIF file format in the Main tab. This can be done either by uploading a solvent-free CIF file of a MOF without partial occupancies or by constructing a CIF file for a hypothetical MOF from linkers and SBU building blocks selected by the user. For the latter option, MOFSimplify uses the Topologically Based Crystal Constructor (ToBaCCo) 3.0 code.[15,79] Prior to assembly, the user must select a compatible linker, SBU, and MOF net combination from dropdown menus. Incompatible combinations are rejected by MOFSimplify.

Once the user selects a MOF for analysis, MOFSimplify can use trained ANN models[57] to make predictions for the structure in the Main tab (Figure 3). MOFSimplify generates RAC features and geometric descriptors of the selected MOF. MOFSimplify then normalizes the descriptors to zero mean and unit variance and uses them as ML model inputs. If the selected MOF is present in the relevant solvent removal stability or thermal stability training data for which a prediction is requested, MOFSimplify returns the ground truth for the selected MOF instead of a prediction. The web server determines the presence or absence of the selected MOF in the training data by comparing RAC and geometric descriptors generated for the selected MOF to the descriptors previously generated for the training data. For predictions, MOFSimplify reports the degree of model confidence through latent space distances for both models and the raw probability output for the solvent removal stability classifier model.[77] The MOFSimplify reports the latent space nearest neighbor (LSNN), which are the MOFs in training data that appear most proximal in the ANN latent space to the loaded MOF for either thermal stability or solvent removal stability. For a good model, a prediction with higher uncertainty is likely to have greater error than a prediction with low uncertainty. Once a prediction is requested and either the ground truth or an ML model prediction is returned, the user can download the descriptors generated for the MOF.

The user can also display and download information about the identified LSNN MOFs, as mentioned in Data Records. This includes structures for LSNNs downloaded as CIF files along with LSNN metadata such as the latent space distance to the selected MOF, DOI of the associated manuscript, and experimentally determined stability can be downloaded in TXT file format. In addition, MOFSimplify allows the user to view a simplified experimental TGA plot for thermal stability ANN LSNNs generated from four TGA trace points or to download the same data in CSV format (see Methods).

If a solvent removal stability prediction is requested, MOFSimplify will either report a ground truth (i.e., stable or unstable) for the selected MOF, or it will display a prediction between 0 (confidently unstable) and 1 (confidently stable) and a sentence reflecting model confidence. If a thermal stability prediction is requested, MOFSimplify reports the prediction/ground truth temperature for the selected MOF ($T_d$*) and the percentile rank of $T_d$* relative to the training data $T_d$ (Figure 3). In addition, MOFSimplify displays the location of $T_d$* relative to the distribution of the thermal ANN training data $T_d$, to give context for $T_d$*.

Further functionality offered by MOFSimplify, both for selected MOFs and LSNN MOFs, includes visualization in the Visualization tab and separation into constituent inorganic SBUs and organic linkers in the Component Analysis tab. MOFSimplify uses both 3Dmol.js[80] and code from the MOFid[53] website for MOF unit cell visualization along with molSimplify to separate MOFs into their constituent parts.[18] MOFSimplify allows the user to filter these MOF components by their atom-weighted molecular graph determinants to isolate unique components as determined by graph connectivity. By default, MOFSimplify does not apply the filter and instead displays all copies identified in the CIF unit cell. MOFSimplify can visualize component structures using 3Dmol.js[80] and display their SMILES codes that are generated with Open Babel.[81,82] The user can download these MOF components as XYZ files (Figure 3).

Additionally, MOFSimplify encourages community engagement by enabling the user to add new MOF data to our database by uploading MOF CIF files and TGA traces in the Data Upload tab. MOFSimplify also lets the user indicate whether they agree with an ANN prediction or curated experimental data and support their position by uploading a TGA trace (Figure 3). These TGA traces will be digitized by us to extract $T_d$ data in a manner consistent with our previous thermal stability data. User input will be used to improve our ANN models through community-based active learning.

As an alternative to using the MOFSimplify web interface, users may also download compiled data for solvent removal stabilities and thermal stabilities in JSON or Excel formats (see Data Records).

**6. Code Availability.**

All scripts used to mine the extant literature corresponding to the CoRE MOF 2019 database are commented and are available on a public GitHub repository at https://github.com/hjkgrp/text_mining_tools. Manuscript copyrights are retained by the publishers, preventing the complete dissemination of full-length articles, but the mined data is provided with an open source CC-BY license and is available on Zenodo at 10.5281/zenodo.5508357 (See Data Records).

The MOFSimplify website is located at https://mofsimplify.mit.edu. The code backend for the MOFSimplify website is available in a public GitHub repository at https://github.com/hjkgrp/MOFSimplify. The repository contains a user manual for the website.

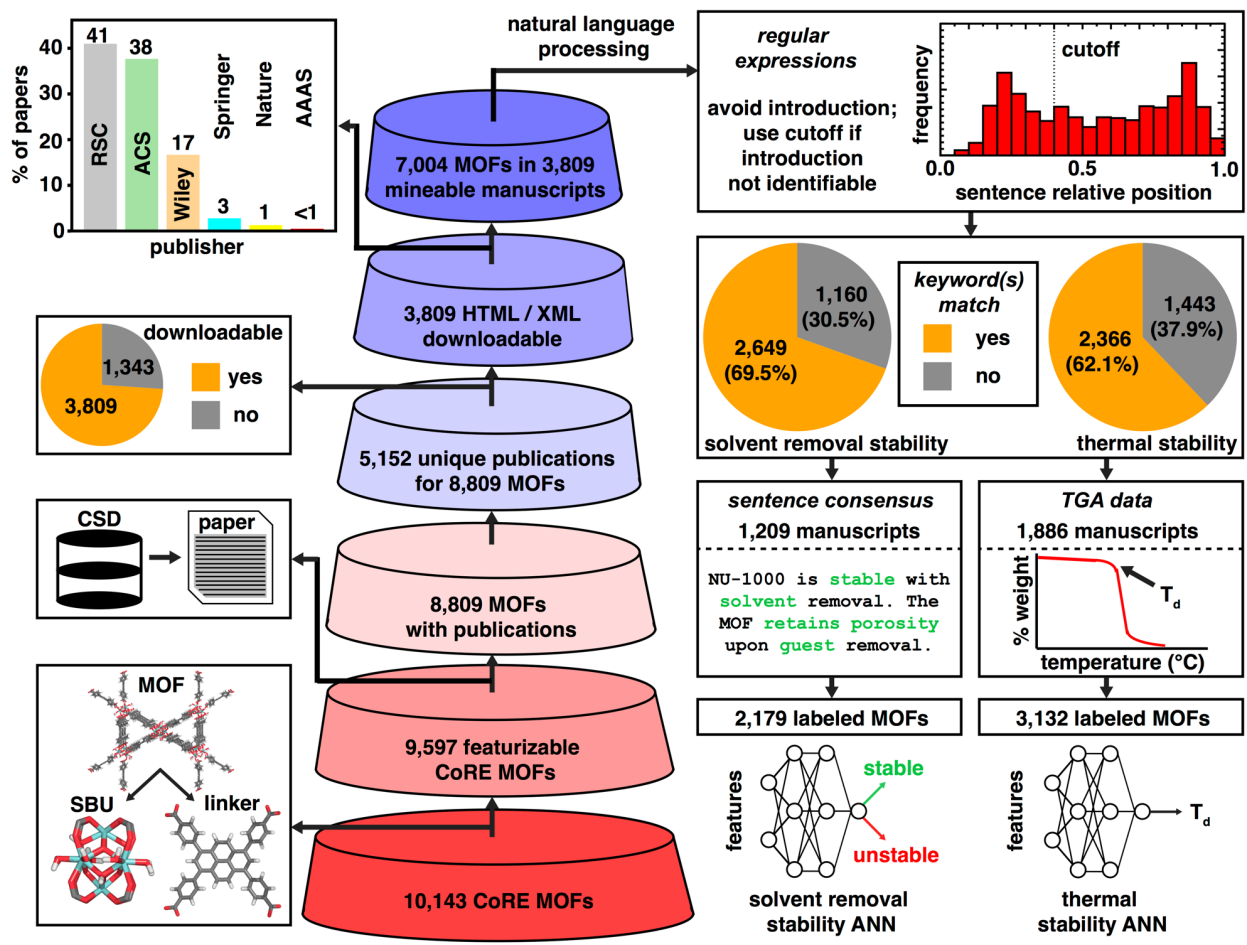

**Figure 1.** Workflows for curating datasets for solvent removal and thermal stability. First, we use sanitized MOFs from published works, filter by structures that can be featurized, obtain manuscripts corresponding to structures, download these manuscripts to prepare them for natural language processing, and finally text mine the manuscripts to identify mentions of solvent removal stability or thermogravimetric analysis data. We identify thermogravimetric analysis traces from manuscripts with thermogravimetric analysis keywords. The two sets of data gathered during this workflow are then used to train machine learning models.

**Table 1.** Keywords used for regular expression searches for solvent removal and thermal stabilities. Stemmed forms of each word were used to identify keywords that have different tenses or forms. We label each word with a category and the type of stability that it identified.

| stemmed keyword | keyword category | words identified | stability type |
|---|---|---|---|
| collaps | collapse | collaps(e/ed/ing) | solvent removal |
| deform | collapse | deform(ed/s/ing/ation) | solvent removal |
| amorph | collapse | amorph(ous/ize) | solvent removal |
| blockage | collapse | blockage | solvent removal |
| degrad | collapse | degrad(e/ed/es/ing/ation) | solvent removal |
| unstable | collapse | unstable | solvent removal |
| instability | collapse | instability | solvent removal |
| destroy | collapse | destroy(ed/s/ing) | solvent removal |
| one step weight | collapse | one(-)step weight loss | solvent removal |
| single step weight | collapse | single(-)step weight loss | solvent removal |
| stable | stable | stable | solvent removal |
| stability | stable | stability | solvent removal |
| preserv | stable | preserv(e/ed/es/ing) | solvent removal |
| crystallinity | stable | crystallinity | solvent removal |
| coordinatively unsaturat | stable | coordinatively unsaturat(ed/ing) | solvent removal |
| porosity | stable | (micro)porosity | solvent removal |
| retain | stable | retain(ed/s/ing) | solvent removal |
| maintain | stable | maintain(ed/s/ing) | solvent removal |
| two step weight | stable | two(-)step weight loss | solvent removal |
| solvent | solvent | solvent(s) | solvent removal |
| guest | solvent | guest(s) | solvent removal |
| desolvat | solvent | desolvat(e/ed/es/ing) | solvent removal |
| remov | solvent | remov(e/ed/es/ing) | solvent removal |
| activat | solvent | activat(e/ed/es/ing) | solvent removal |
| evacuat | solvent | evacuat(e/ed/es/ing) | solvent removal |
| dehydrat | solvent | dehydrat(e/ed/es/ing) | solvent removal |
| eliminat | solvent | eliminat(e/ed/es/ing) | solvent removal |
| water, H2O | solvent | water, H2O | solvent removal |
| DMF, formamide | solvent | DMF, formamide | solvent removal |
| DMA, methylamine, diamine | solvent | DMA, methylamine, diamine | solvent removal |
| EtOH, MeOH, ethanol, methanol | solvent | EtOH, MeOH, ethanol, methanol | solvent removal |
| pyrrolidone | solvent | pyrrolidone | solvent removal |
| TG | thermal | TG(A) | thermal |
| thermogravimetric | thermal | thermogravimetric analysis | thermal |
| thermal gravimetric | thermal | thermal(-)gravimetric analysis | thermal |

**Table 2**. Description of revised autocorrelation (RAC) features with start/scope, operation performed, count of features removed, and total feature count. Five heuristic atom-wise quantities are used to perform all product and difference operations: nuclear charge (Z), electronegativity ($\chi$), topology (T), identity (I), and covalent radius (S). MOF RACs contain five possible starts and two possible scopes: metal-centered (mc) start, linker coordinating atom centered (lc) start, functional group centered (func) start, every atom (full) start, all atom in primitive cell (all) scope, or all atom in linker (linker) scope. All starts, scopes and operations use bond depths of 0, 1, 2, and 3 to generate autocorrelations (for a total of 20 possible features for each scope). Cases that are invariant across all MOFs are listed in the "features removed" column. RAC features are given using the notation: <operation/start>-<atomic property>-<depth>-<scope>. RAC features using products are denoted by their start (e.g. mc), and those using differences contain a "D" prefix with a subscripted start (e.g. $D_{mc}$).

| start | scope | operation | features removed | feature count |
|---|---|---|---|---|
| mc | all | product | 1 (mc-I-0-all) | 19 |
| mc | all | difference | 8 ($D_{mc}$-I-0-all, $D_{mc}$-I-1-all, $D_{mc}$-I-2-all, $D_{mc}$-I-3-all, $D_{mc}$-S-0-all, $D_{mc}$-T-0-all, $D_{mc}$-Z-0-all, $D_{mc}$-$\chi$-0-all) | 12 |
| lc | linker | product | 1 (lc-I-0-linker) | 19 |
| lc | linker | difference | 8 ($D_{lc}$-I-0-linker, $D_{lc}$-I-1-linker, $D_{lc}$-I-2-linker, $D_{lc}$-I-3-linker, $D_{lc}$-S-0-linker, $D_{lc}$-T-0-linker, $D_{lc}$-Z-0-linker, $D_{lc}$-$\chi$-0-linker) | 12 |
| func | linker | product | 0 | 20 |
| func | linker | difference | 8 ($D_{func}$-I-0-linker, $D_{func}$-I-1-linker, $D_{func}$-I-2-linker, $D_{func}$-I-3-linker, $D_{func}$-S-0-linker, $D_{func}$-T-0-linker, $D_{func}$-Z-0-linker, $D_{func}$-$\chi$-0-linker) | 12 |
| full | all | product | 0 | 20 |
| full | linker | product | 0 | 20 |
|  |  |  | 26 | 134 |

**Table 3.** Description of geometric features generated by Zeo++ with definitions and units.

| variable name | explanation | units |
|:---:|:---:|:---:|
| $D_f$ | maximum free sphere | Å |
| $D_i$ | maximum included sphere | Å |
| $D_{if}$ | maximum included sphere in the free sphere path | Å |
| GPOAV | gravimetric pore accessible volume | $cm^3/g$ |
| GPONAV | gravimetric pore non-accessible volume | $cm^3/g$ |
| GPOV | gravimetric pore volume | $cm^3/g$ |
| GSA | gravimetric surface area | $m^2/g$ |
| POAV | pore accessible volume | $Å^3$ |
| PONAV | pore non-accessible volume | $Å^3$ |
| POAVF | pore accessible volume fraction | unitless |
| PONAVF | pore non-accessible volume fraction | unitless |
| VPOV | volumetric pore volume | $cm^3/cm^3$ |
| VSA | volumetric surface area | $m^2/cm^3$ |
| $\rho$ | crystal density | $g/cm^3$ |

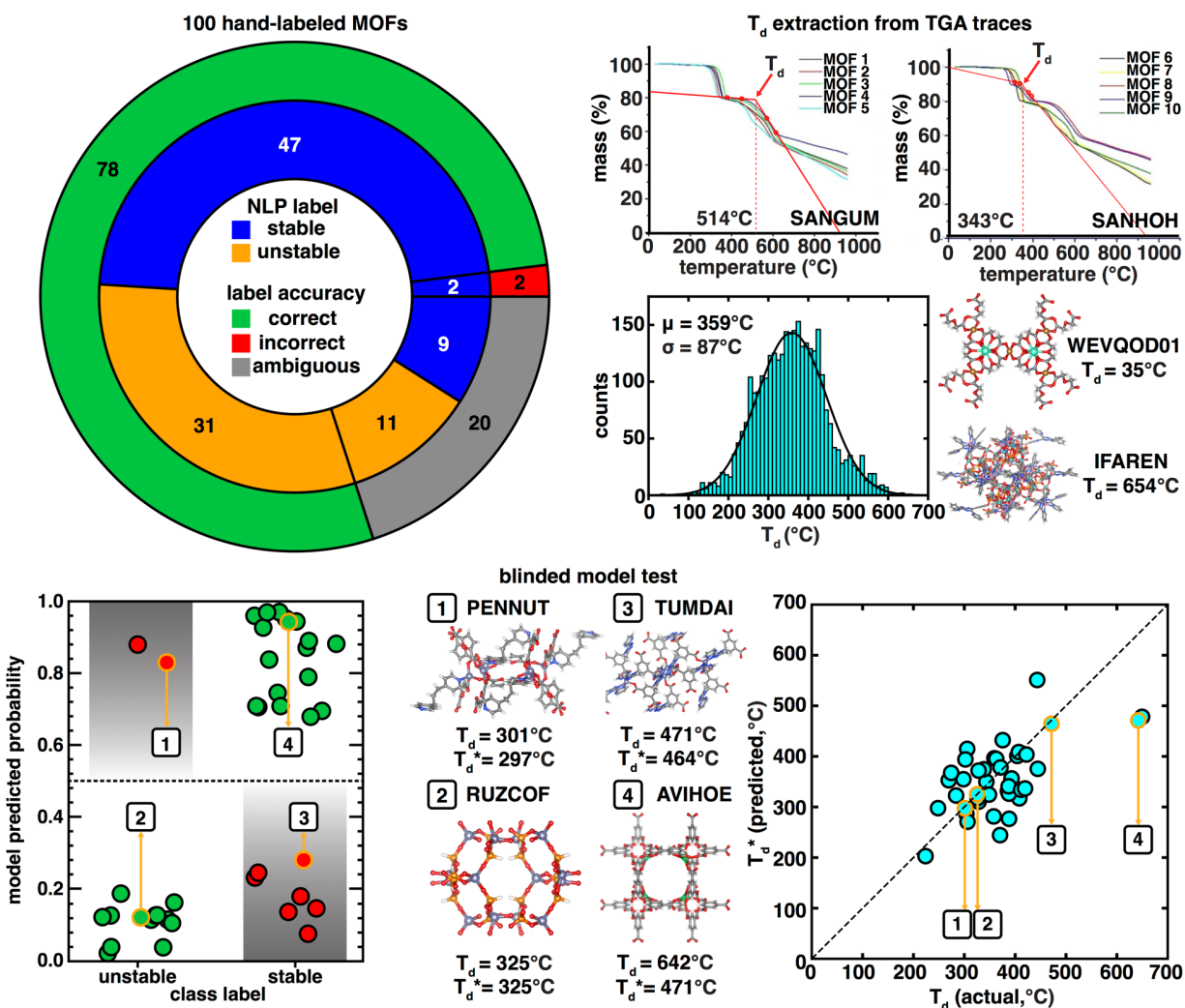

**Figure 2.** Validation of the solvent removal and thermal stability data sets. Comparison of NLP-assigned labels to hand-assigned labels over a 100 MOF subset (top left), with stable NLP-assigned labels in blue, and unstable labels in orange. Cases that were correctly assigned are shown with a green outer ring, those that were incorrect are shown with a red outer ring, and ambiguous cases are shown with a gray outer ring. Assignment of $T_d$ from TGA traces (top right, TGA traces adapted from ref. [78]) shown for two MOFs (SANGUM and SANHOH), with $T_d$ values denoted inset. The distribution of $T_d$ over the full thermal stability dataset is shown, with the MOF containing the lowest (WEVQOD01) and highest (IFAREN) thermal decomposition temperatures shown inset. A blinded test over 40 MOFs (bottom) demonstrates the accuracy of literature-trained ANN ML models for solvent removal stability prediction (bottom left) and thermal stability prediction (bottom right), with four representative examples (PENNUT, RUZCOF, TUMDAI, and AVIHOE) shown inset.

**Figure 3.** Sections of the MOFSimplify web interface. (A) Interface for selecting a MOF for analysis and predicting properties of the selected MOF using ANNs trained on experimental data mined from the literature. The default MOF loaded upon selecting "Example MOF" is HKUST-1, a well-studied MOF.[83] (B) The feedback interface for evaluating model predictions. (C) The interface listing similar (i.e., LSNN) MOFs to the selected MOF as determined by the ANNs. (D) Visualization of the selected MOF's components. (E) Visualization of the selected MOF's unit cell.

## ACKNOWLEDGMENT

The authors acknowledge support by DARPA (grant number D18AP00039) for the text extraction, machine learning, and website development efforts (to A.N., G.T., C.D., and N.A.), and some of the database curation efforts were supported by the Office of Naval Research under grant number N00014-20-1-2150 (to C.D. and H.J.K). This work is also partially supported as part of the Inorganometallic Catalysis Design Center, an Energy Frontier Research Center funded by the U.S. Department of Energy, Office of Science, Basic Energy Sciences under Award DE-SC0012702 (to A.N. and G.T.). This work was also partially supported by a National Science Foundation Graduate Research Fellowship under Grant #1122374 (to A.N., N.A., and D.W.K.). H.J.K. holds a Career Award at the Scientific Interface from the Burroughs Wellcome Fund and an AAAS Marion Milligan Mason Award, which supported this work. The authors thank Adam H. Steeves and Akash Bajaj for providing a critical reading of the manuscript. The authors also acknowledge helpful conversations with members of the Elsa Olivetti lab.

## AUTHOR CONTRIBUTIONS

A.N. curated the data and performed analyses on data validity. A.N. and C.D. worked on the ML model training. G.T. constructed the MOFSimplify website, with contributions from N.A., A.N., C.D., and D.W.K.. A.N., G. T., and H.J.K. wrote the manuscript. All authors contributed to revising the manuscript.

## CORRESPONDING AUTHOR

Correspondence to Heather J. Kulik, email: hjkulik@mit.edu, phone: 617-253-4584

COMPETING INTERESTS

The authors declare no competing financial interests.